\documentclass[12pt,preprint]{emulateapj}
\usepackage{natbib}

\slugcomment{To appear in ApJ}

\shorttitle{SONYC: NGC1333}
\shortauthors{Scholz et al.}

\begin{document}
\bibliographystyle{apj}


\title{Substellar Objects in Nearby Young Clusters (SONYC) VI: \\
The planetary-mass domain of NGC1333}


\author{Alexander Scholz\altaffilmark{1}, Ray Jayawardhana\altaffilmark{2,**}, Koraljka Muzic\altaffilmark{2}, 
Vincent Geers\altaffilmark{3}, Motohide Tamura\altaffilmark{4}, Ichi Tanaka\altaffilmark{5}}

\email{aleks@cp.dias.ie}

\altaffiltext{1}{School of Cosmic Physics, Dublin Institute for Advanced Studies, 31 Fitzwilliam Place,
Dublin 2, Ireland}
\altaffiltext{2}{Department of Astronomy \& Astrophysics, University of Toronto, 50 St. George Street, Toronto, 
ON M5S 3H4, Canada}
\altaffiltext{3}{Institute for Astronomy, ETH Zurich, Wolfgang-Pauli-Strasse 27, 8093 Zurich, Switzerland}
\altaffiltext{4}{National Astronomical Observatory, Osawa 2-21-2, Mitaka, Tokyo 181, Japan}
\altaffiltext{4}{Subaru Telescope, National Astronomical Observatory of Japan, 650 North A'ohoku Place, Hilo, HI 96720, USA}
\altaffiltext{**}{Principal Investigator of SONYC}

\begin{abstract}
Within the SONYC -- {\it Substellar Objects in Nearby Young Clusters} -- survey, we investigate 
the frequency of free-floating planetary-mass objects (planemos) in the young cluster NGC~1333. 
Building upon our extensive previous work, we present spectra for 12 of the faintest candidates 
from our deep multi-band imaging, plus seven random objects in the same fields, using MOIRCS on 
Subaru. We confirm seven new sources as young very low mass objects (VLMOs), with T$_{\mathrm{eff}}$ of
2400--3100\,K and mid-M to early-L spectral types. These objects add to the growing census of VLMOs 
in NGC1333, now totaling 58. Three confirmed objects (one found in this study) have masses below 
15\,M$_{\mathrm{Jup}}$, according to evolutionary models, thus are likely planemos. We estimate 
the total planemo population with 5-15\,M$_{\mathrm{Jup}}$ in NGC1333 is $\la 8$. The mass spectrum 
in this cluster is well approximated by $dN/dM \propto M^{-\alpha}$, with a single value of $\alpha = 0.6\pm 0.1$ for 
$M<0.6\,M_{\odot}$, consistent with other nearby star forming regions, and requires $\alpha \la 0.6$ 
in the planemo domain. Our results in NGC1333, as well as findings in several other clusters by 
ourselves and others, confirm that the star formation process extends into the planetary-mass domain, 
at least down to $6\,M_{\mathrm{Jup}}$. However, given that planemos are 20-50 times less numerous than stars, 
their contribution to the object number and mass budget in young clusters is negligible. Our findings 
disagree strongly with the recent claim from a microlensing study that free-floating planetary-mass 
objects are twice as common as stars -- if the microlensing result is confirmed, those isolated 
Jupiter-mass objects must have a different origin from brown dwarfs and planemos observed in young 
clusters.
\end{abstract}


\section{Introduction}
\label{s1}

The frequency of free-floating objects with masses below the Deuterium burning limit ($<0.015\,M_{\odot}$) --
in the following called 'planemos', short for planetary-mass objects -- is a subject of debate in the literature.
Deep surveys in nearby star forming regions indicate the presence of at least some planemos 
\citep[e.g.][]{1998Sci...282.1095T,2000Sci...290..103Z,2000MNRAS.314..858L,2006ApJ...647L.167J,2008ApJ...675.1375L,2010ApJ...709L.158M}, 
but they yield conflicting results regarding their numbers. On the other hand, a recent micro-lensing study 
\citep{2011Natur.473..349S} reports the presence of a large population of Jupiter-mass objects in the 
field ('almost twice as common as stars'), which are either without host star or on very wide orbits. 
Direct detections of field objects in this mass domain have been reported as well 
\citep[e.g.][]{2011ApJ...740..108L,2012ApJ...748...74L}. So far, it is unclear whether the mass function 
declines in the planemo domain or not. Young planemos are also useful as benchmark objects for testing 
models for atmospheres and evolution in a new mass and age domain.

\begin{figure}
\center
\includegraphics[width=9cm]{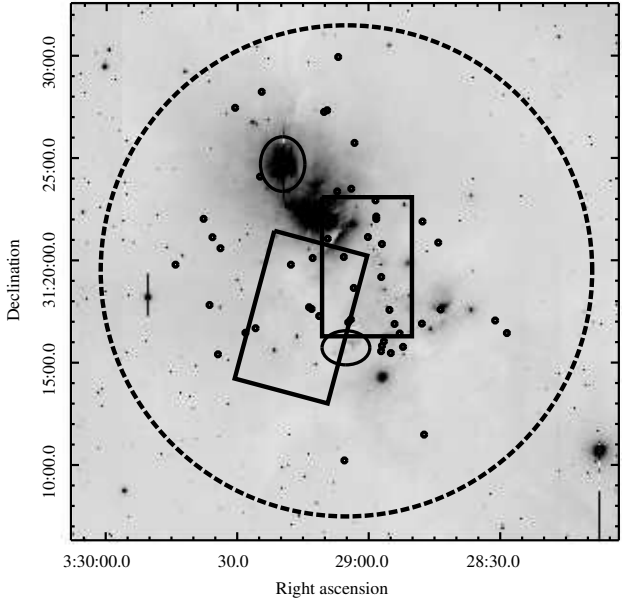} 
\caption{Subaru/Suprime-Cam i-band image for our target region NGC 1333 in the Perseus star-forming region. This image
is identical to Fig. 1 in \citet{2012ApJ...744....6S}, but we have overplotted with solid lines the two MOIRCS fields observed
for this paper. Marked are the cluster radius with a large dashed circle, the two well-known objects BD+30 549 (north) 
and HH 7-11 (south) with small ellipses, and the known population of very low mass members (Scholz et al. 2009) with small 
circles. The reflection nebula NGC 1333 is slightly north of the image center. Coordinates are J2000. 
\label{f0}}
\end{figure}

One of the goals of our SONYC survey (short for Substellar Objects in Nearby Young Clusters) is to probe the
planemo domain in several nearby star forming regions, using broadband imaging surveys followed by 
spectroscopic verification of the candidates. One of our targets is the young cluster NGC1333, a $\sim 1$\,Myr 
old compact star forming region in the Perseus OB association (see Fig. \ref{f0}). In this cluster we found a rich 
substellar population, with about 30-40 brown dwarfs \citep[][hereafter SONYC-I]{2009ApJ...702..805S}. In our most recent paper 
on this cluster we also report the discovery of a handful of objects with estimated masses below 0.02$\,M_{\odot}$, 
one of them at roughly 0.006$\,M{\odot}$ and thus firmly in the planemo regime 
\citep[][hereafter SONYC-IV]{2012ApJ...744....6S}.

Thus far, the number of planemos in this cluster seems low compared with other regions. This, however, is a preliminary 
result, for two reasons: 1) Our spectroscopic follow-up was not sufficiently complete in the planemo domain. 2) Our 
optical survey might not be sufficiently deep in some of the parts of NGC1333 subject to high attenuation
due to reddening. Here we aim to resolve these issues through additional spectroscopy of faint candidates and a 
cumulative analysis, thus providing more definitive constraints on the planemo population of this cluster.

\section{Target selection}
\label{s2}

The goal of this paper is to probe specifically the planemo regime in NGC1333 and to address the potential biases outlined
in Sect. \ref{s1}. To that end, we selected three samples of objects for follow-up spectroscopy, in total 19 objects,
which are listed in Tables \ref{t1} and \ref{t2}. 

\begin{figure}
\center
\includegraphics[width=6cm,angle=-90]{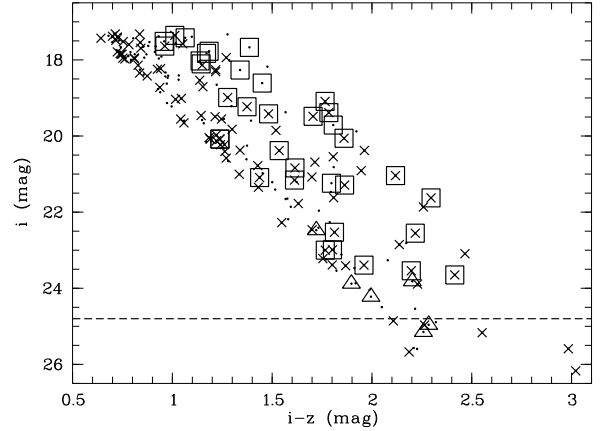} 
\caption{Colour-magnitude diagram for the iz candidates (dots), originally identified in SONYC-I. The figure is
identical to Fig. 7 in SONYC-IV, but we have also marked the candidates targeted in the new campaign with triangles.
Crosses are objects for which we have already obtained spectra in previous campaigns. Confirmed very low mass objects are marked 
with squares. The horizontal dashed line shows the completeness limit of the photometric survey as estimated in SONYC-I.
\label{f2}}
\end{figure}

{\it 1. IZ candidates:} Here we started with our primary set of 196 candidates identified based on the (i,i-z) colour-magnitude
diagram (SONYC-I). This diagram is shown in Fig. \ref{f2}; objects for which we previously obtained spectra are
marked with crosses, confirmed very low mass objects with squares. A subsample of 22 objects with $i>22$ has not yet been
verified spectroscopically. 20 of these objects are within 0.25\,deg of the cluster center, which is the area where all confirmed
very low mass members are located. From these 20, we select 6 for follow-up spectroscopy, marked with triangles in Fig. \ref{f2}.
As can be seen in the figure, these candidates cover i-band magnitudes from 22.5 to 25, extending beyond the
nominal completeness limit of the photometric survey (dashed line). Including this new selection, 27 out of 43 IZ candidates with 
$i>22$ and 9 out of 15 with $i>24$ have been observed spectroscopically. The lowest mass confirmed member of the cluster, 
SONYC-NGC1333-36, with an estimated mass of 0.006$\,M_{\odot}$ appears in this plot at $i=23.39$ and $i-z=1.96$.

\begin{figure}
\center
\includegraphics[width=6cm,angle=-90]{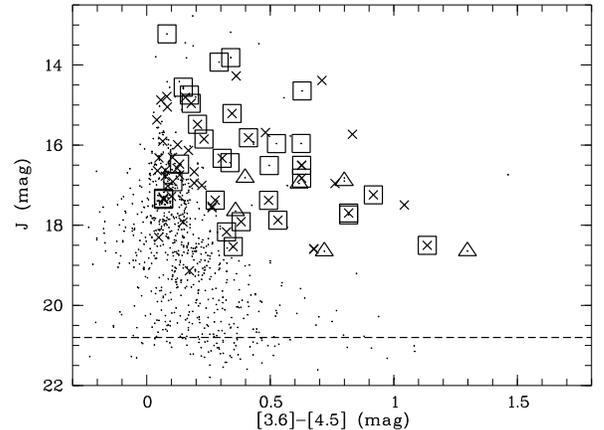} 
\caption{Colour-magnitude diagram for the JKS candidates (dots), see text for details. Confirmed very low mass objects are marked 
with squares, crosses are objects for which we have already obtained spectra in previous campaigns. Objects targeted in this new 
campaign are marked with triangles. The completeness limit of the J-band photometry is marked with a dashed line.
\label{f3}}
\end{figure}

{\it 2. JKS candidates:} We correlated the deep JK catalogue presented in SONYC-I with the 'HREL' catalogue
from the Spitzer 'Cores to Disks' (C2D) legacy program \citep{2009ApJS..181..321E}. We accepted a source if it has a Spitzer 
counterpart within 1" and uncertainties below 0.2\,mag in the first two IRAC bands at 3.6 and 4.5$\,\mu m$ (hereafter I1 and 
I2). This yields a catalogue of 824 objects (called the JKS catalogue). The (J,I1-I2) colour-magnitude diagram for this 
sample is plotted in Fig. \ref{f3}. The figure illustrates that this catalogue is significantly deeper than the IZ selection: 
the faintest IZ candidates have $J\sim 19$, whereas the JKS catalogue is complete down to $J=20.8$. From this catalogue, we 
only consider the subsample with $I1-I2>0.3$. This cut-off was introduced to avoid the bulk of the datapoints in the diagram, 
for which a spectroscopic follow-up is not practical; in Sect. \ref{s52} we will discuss the consequences of this cut on the 
completeness of the survey. There are only 17 objects without spectroscopic verification with $16<J<19$, $I1-I2>0.3$ and 
(as above) a distance from the cluster center below 0.25\,deg. Out of these 17, we selected 6 for follow-up spectroscopy, 
which are marked with triangles. The aforementioned SONYC-NGC1333-36 appears in this plot at $J=18.53$ and $I1-I2=0.35$. 
Note that spectroscopic verification for objects beyond $J=19$ is not feasible with the currently available facilities.

{\it 3. Random candidates:} After selecting the IZ and JKS candidates, we decided to observe two MOIRCS fields (see Sect. \ref{s3}
and Fig. \ref{f0}), which cover a large fraction of the samples. To prepare the spectroscopy run, we took a new set of deep 
K-band images for these two fields. From these images, we selected 7 more sources for spectroscopy, for which we had no prior
information about their colours. The only constraint for these random sources was the outline of the slit mask.

We note that most of the selected objects are listed as candidates from photometric surveys in the literature,
in papers by \citet{1994A&AS..106..165A,1996AJ....111.1964L,2004AJ....127.1131W,2008AJ....136.1372O,2008ApJ...674..336G}.

Two caveats need to be pointed out with regard to the selection of candidates from colour-magnitude diagrams, as done here for the 
IZ and JKS candidates. 

First, the completeness of the survey depends strongly on the definition of the colour and magnitude criteria used to select 
the candidates for follow-up spectroscopy. For the IZ sample, we defined the selection box based on the position of previously 
identified brown dwarfs and extend it at the faint end -- we are simply probing the extension of the colour space 
occupied by the known brown dwarfs towards fainter magnitudes. A similar strategy was used for the JKS sample, but as argued
above, we need to introduce an additional colour cut (see also the discussion Sect. \ref{s52}). In the two samples, the colour 
cuts are independent of evolutionary models.

Second, the photometric uncertainties may cause us to miss some brown dwarfs. For the iz selection, 
this is not a major problem: The median error (excluding systematics) for $i>22$ is 0.05\,mag, and 95\% of the errors
are below 0.1\,mag. The width of the selection box (1\,mag in (i-z) colour) is significantly larger the errors. In addition, 
most of the confirmed objects are 0.1-0.2\,mag to the right side of the left boundary of the box. Assuming that these objects 
show a reliable representation of the cluster isochrone in NGC1333, even large uncertainties are unlikely to move the datapoints 
of substellar members out of our selection box. Note that the right side of our selection criterion was only formal, no objects 
in our survey had colours redder than this boundary.

In the JKS sample, the median error in the (I1-I2) colour is 0.1\,mag for $16<J<19$. For $18<J<19$, i.e. at the faint end of our
candidate selection, the errors are $\le 0.22$\,mag. These errors do not depend significantly on the (I1-I2) colour. The 
expected (I1-I2) colours of brown dwarfs and planemos are close to the bulk of the datapoints in this diagram (see Sect. 
\ref{s52}), i.e. we may lose objects on the left side of our colour cut due to large photometric errors. 

\section{Observations and data reduction}
\label{s3}

New spectra for candidate planemo members of NGC1333 were taken using the near-infrared spectrograph MOIRCS at the Subaru telescope
in multi-object (MOS) mode. Since our targets are faint and require long integration times, MOS is the most practical way of
conducting a spectroscopic survey. After creating the IZ and JKS candidate samples, we looked for the optimum way to obtain 
spectra for as many as possible of these objects. We found that two MOIRCS fields arranged as shown in Fig. \ref{f0} can cover 12 
of these candidates (6 IZ and 6 JKS, as discussed in Sect. \ref{s2}). 

K-band pre-imaging was carried out for these fields in September 2011, with excellent seeing (0.3") and under good conditions.
The pre-images were reduced and calibrated using 2MASS photometry. Aperture magnitudes were measured for all our candidates, to
complement and verify the available photometry.
These pre-images are used to define the masks for the spectroscopy run, which was carried out in the night of Nov 14-15. For most of 
this night, we observed under stable conditions and with seeing
in the range of 0.5". At 2am local time, about 2 hours before the end of the visibility for our target region, the humidity
increased above 80\% forcing us to close the dome. In total, we spent 3:20\,h telescope time on mask 1 and 3:10\,h on mask 2.
At the beginning of the night and between mask 1 and mask 2 we observed 3 standard stars with spectral type around A0 intended to
be used for telluric calibration.

We used our two MOS masks, the grism HK500 with a nominal resolution of 600-800 and a wavelength coverage from 1.3 to 2.4$\,\mu m$, 
an order sorting filter, and slits with a length of 11.7" and a width of 1.17". For each 
mask we obtained a series of individual exposures of at most 5\,min. Due to an incident with the spillage of cooling liquid earlier
in the year (July 2 2011), 
the autoguider system was not operational at the time of the observations. Therefore, we checked the position of the mask 
alignment stars between individual exposures and re-aligned the mask if the offset was more than the typical seeing in the direction 
perpendicular to the slit. If the offsets were large, we reduced the integration time to be able to track the sources reliably.
The mask position was typically stable over several minutes, the integration times for individual exposures were 240\,sec for 
mask 1 and 80-300\,sec for mask 2. The total on-source time was 88\,min for mask 1 and 60\,min for mask 2. Between consecutive
exposures, the telescope was nodded along the slits by 2".

We reduced the MOIRCS spectra using the recipes described in detail in SONYC-I. This includes the a) subtraction of
consecutive exposures that are offset along the slit, b) division by a normalised domeflat calculated from lamp-on
and lamp-off exposures, and c) the stacking of the individual frames for each mask. For the last step, we measured the 
position of one of the sources in the slit for each individual frame and shifted the frames accordingly, to account for small
offsets.

Spectra were extracted using {\it apall} in IRAF. This includes aperture definition, background subtraction, and trace fitting.
We opted for a full aperture integration across the profile (as opposed to optimum extraction), because many of our sources are faint
and lack a well-defined profile. The same procedure was carried out for the standard stars. The wavelength solution was determined
from $\sim 10$ telluric lines in the raw spectra. For each target, the two resulting spectra, corresponding to the two positions
in the slit, were coadded. The final spectra are re-binned to 40\AA~per wavelength element.

The three standards HIP6855, HIP9196, and HIP21115 have been observed at airmasses of 1.46, 1.64 and 1.10, while the average airmass 
for the science exposures was 1.30 for mask 1 and 1.06 for mask 2. The spectrum for HIP6855 has a significantly lower flux level in 
the K-band (by 10-20\%) than the two other stars and was not used. The scaled spectra for HIP9196 and HIP21115, however, are 
very similar with small differences in the range of 1-2\%, well-explained by the differences in airmass. This is consistent with 
an extinction coefficient difference (mag/airmass) in the range of 0.01 between H- and K-band, which is plausible for Mauna Kea.

Because HIP21115 was observed between the two masks and thus should give the best approximation to the conditions during the science 
exposures, we used its spectrum for the calibration. All science spectra were divided by the quotient between the telluric 
standard spectrum and the tabulated spectrum of an A0 star \citep{1998PASP..110..863P}.\footnote{Although HIP21115 is a B9V star,
we used the tabulated A0 spectrum because the Pickles et al. spectrum for spectral type A0 has much better resolution. The
difference in the flux levels between A0 and B9 do not matter in the H- and K-band for our purposes.}

\section{Spectral analysis} 
\label{s4}

The 19 available spectra were analysed using the recipes described in detail in SONYC-IV. In the following, we give
a brief summary of the four steps and describe the results. 

\begin{figure}
\center
\includegraphics[width=8cm,angle=0]{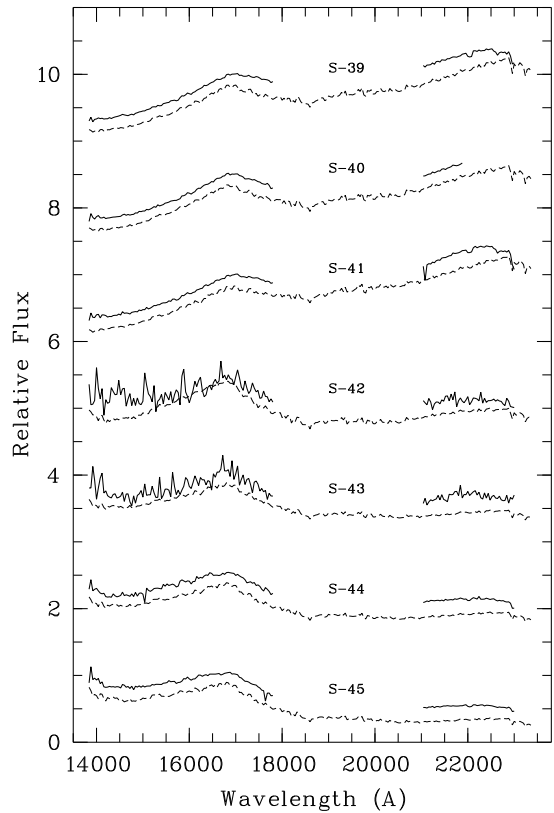} 
\caption{Near-infrared spectra of the 7 newly identified very low mass members of NGC1333 (solid lines). The
dashed lines show the best-fitting, reddened model spectra with the parameters as listed in Table \ref{t1}.
The spectra are offset on the y-axis by constants for clarity.
\label{f6}}
\end{figure}

{\it 1. Selection of young very low mass objects:} 7 spectra show the characteristic spectral signature of young very low mass
objects. Specifically, they exhibit the sharp 'H-band peak' at 1.68$\,\mu m$ caused by broad water absorption features at 
1.3-1.5 and 1.75-2.05\,$\mu m$ \citep{2005ApJ...623.1115C}, which is an indication for low-gravity \citep{2006ApJ...653L..61B}.
These 7 sources are considered to be very low mass objects. Given the compact nature of NGC1333 and the low space densities for 
these objects \citep{2008A&A...488..181C}, contamination by foreground or background objects is negligible. Thus, these 7 objects 
are most likely young members of NGC1333. Their spectra are shown in Fig. \ref{f6}; their properties listed in Table \ref{t1}. 
The remaining 12 spectra are either featureless or show only little structure in the H-band. They are listed in Table \ref{t2}.

\begin{deluxetable*}{cllccccccl}
\tabletypesize{\scriptsize}
\tablecaption{New very low mass members of NGC1333
\label{t1}}
\tablewidth{0pt}
\tablehead{
\colhead{ID} & \colhead{$\alpha$(J2000)} & \colhead{$\delta$(J2000)} & \colhead{J (mag)} & \colhead{K (mag)} & \colhead{SpT} &
\colhead{$A_V$ (phot)} & \colhead{$A_V$ (spec)} & \colhead{$T_{\mathrm{eff}}$} & \colhead{Comments}}
\tablecolumns{10}
\startdata
39 & 03:28:54.96 & +31:18:15.3 & 16.881 & 13.155 & M8.5 & 15 & 17 & 2900 & IZ  \\ 
40 & 03:29:02.80 & +31:22:17.3 & 16.959 & 13.423 & M9.3 & 14 & 15 & 2700 & JKS \\ 
41 & 03:29:04.63 & +31:20:28.9 & 17.647 & 13.892 & M7.6 & 15 & 18 & 3100 & JKS \\ 
42 & 03:29:13.85 & +31:18:05.7 & 19.252 & 17.540 & ~L1  &  4 & 4  & 2400 & IZ\tablenotemark{a} \\ 
43 & 03:29:21.94 & +31:18:29.2 & 18.728 & 17.128 & M8.1 &  3 & 5  & 2900 & IZ  \\ 
44 & 03:28:57.04 & +31:16:48.7 & --     & 15.59  & M8.6 &    & 4  & 2800 & random \\ 
45 & 03:29:13.04 & +31:17:38.3 & 15.23  & 14.16  & M8.1 &  0 & 2  & 2900 & random \\ 
\enddata
\tablenotetext{a}{Fit manually adjusted to account for poor spectral quality}
\end{deluxetable*}


\begin{deluxetable}{lll}
\tabletypesize{\scriptsize}
\tablecaption{Objects ruled out as very low mass members of NGC1333
\label{t2}}
\tablewidth{0pt}
\tablehead{
\colhead{$\alpha$(J2000)} & \colhead{$\delta$(J2000)} & \colhead{Comments}}
\tablecolumns{3}
\startdata 
03:29:04.31 & +31:19:06.4 & JKS, $T\sim3500$           \\ 
03:28:59.33 & +31 16 31.5 & IZ,  $T>3500$              \\ 
03:29:01.88 & +31:16:53.4 & JKS, $T\sim3500$           \\ 
03:29:02.69 & +31 19 05.6 & IZ,  $T\sim3500$           \\ 
03:28:58.42 & +31:22:17.6 & JKS, featureless           \\ 
03:29:05.67 & +31:21:33.9 & JKS, $T\sim3500$           \\ 
03 29 13.05 & +31 16 39.8 & IZ,  $T>3500$              \\ 
03:29:00.02 & +31:17:13.4 & random, $T>3500$           \\ 
03:28:56.31 & +31:20:35.2 & random, featureless        \\ 
03:29:06.45 & +31:16:52.0 & random, featureless        \\ 
03:29:16.26 & +31:18:06.6 & random, featureless        \\ 
03:29:21.88 & +31:17:40.7 & random, featureless        \\ 
\enddata
\end{deluxetable}

{\it 2. Spectral types:} For the 7 objects from Table \ref{t1} we measured spectral types using the H-peak index
(hereafter HPI)
introduced in SONYC-IV, after dereddening the spectra using the spectroscopic $A_V$ (see below).
As discussed in the previous paper, other near-infrared indices suggested in the literature are not suitable for 
our type of data. The HPI calculates the flux ratio between the wavelength intervals 1.68 and 1.50$\,\mu m$ and 
is usable for $>$M6 spectral types. Using the calibration given in SONYC-IV, all 7 objects identified 
here have spectral types of M7 or later (see Table \ref{t1}), the latest type is $\sim$L1 (albeit with a noisy
spectrum). The typical uncertainty is one subtype.

{\it 3. Effective temperatures:} We calculate a grid of model spectra for effective temperatures from 1500 to
3900\,K (in steps of 100\,K) and extinctions $A_V$ from 0 to 30\,mag. This grid is based on the DUSTY models
by \citet{2001ApJ...556..357A}. For all objects with any indication of structure in the H-band, we searched for the best 
matching model spectrum. This was done in the following way: For each model we subtracted observed spectrum from model 
spectrum, squared this quantity, divided by the model spectrum, and averaged the residual over the wavelength domain 
1.38-1.78 and 2.10-2.32$\,\mu m$. We selected the model spectrum for which this quantity is minimal, and adopted its 
parameters as best fit (see Table \ref{t1}).
The typical uncertainty is $\pm 200$\,K in $T_{\mathrm{eff}}$ and $\pm 1$\,mag in $A_V$. Whenever possible we also
calculated $A_V$ from the J- and K-band photometry. The results are given in Table \ref{t1}. The 7 very low mass
objects have effective temperatures between 2400 and 3100\,K. The remaining sources with recognisable structure
in the H-band have temperature around or above 3500\,K (Table \ref{t2}). The object SONYC-NGC1333-42, the 
coolest source in our sample, has a noisy spectrum; therefore we manually chose the best-fitting model. The
best-fitting reddened models are overplotted in Fig. \ref{f6}.

Two of the objects in Table \ref{t1} were selected randomly. One of them (SONYC-NGC1333-45) turns out to be on our 
initial list of IZ candidates but much brighter than the cutoff used for this paper ($i=18.609$ vs. $i=22$). It was 
previously confirmed by \citet{2009AJ....137.4777W}. They find a spectral type of M8 and a temperature of 2700\,K; our 
analysis yields consistent results (M8 and 2900\,K). The second random source, SONYC-NGC1333-44, has been listed 
before as a photometric candidate \citep{1994A&AS..106..165A,2008AJ....136.1372O}. It was not found in our IZ survey 
because it sits very close to a blooming spike caused by a bright star.

{\it 4. Sanity checks:} As in our previous papers, we additionally compared the spectra to literature spectra of 
other brown dwarfs. This is particularly important to make sure that our results are not severely affected by
the often low signal-to-noise ratio in our data. For comparison we use the spectra for the three young objects 2M1207\,A, 
DH\,Tau\,B, and 2M1207\,B, published by \citet{2012arXiv1201.3921P} in a detailed comparison with model spectra. For
all three they find a satisfying match with different types of model spectra and consistent temperatures for a
variety of spectral ranges and model spectra.
According to their analysis, these objects have effective temperatures of 3100, 2600, and 1600\,K, when HK spectra
are compared with DUSTY models, which is comparable to the way we have determined temperatures. With our method,
we derive $T_{\mathrm{eff}}$ of 3200, 2800, and 1700\,K for the same data, consistent with the published values.

\begin{figure}
\center
\includegraphics[width=7cm,angle=0]{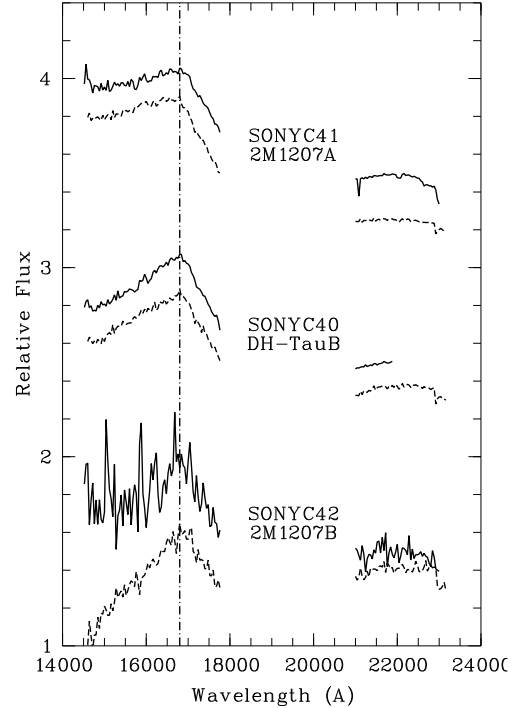} 
\caption{Three spectra for newly identified very low mass members of NGC1333 from Table \ref{t1} (solid
lines) in comparison with published spectra for brown dwarfs from \citet{2012arXiv1201.3921P} with similar 
effective temperature (dashed lines). The spectra are offset on the y-axis by constants for clarity.
\label{f10}}
\end{figure}

In Fig. \ref{f10} we plot their spectra in comparison with three of our newly identified objects in NGC1333 with similar 
temperatures: SONYC-NGC1333-41, -40, and -42 with $T_{\mathrm{eff}}$ of 3100, 2700 and 2400\,K. The SONYC spectra were 
dereddened using the spectroscopic $A_V$ given in Table \ref{t1}. From Fig. \ref{f10} it seems plausible that the two 
pairs SONYC-NGC1333-41/2M1207\,A and SONYC-NGC1333-40/DH\,Tau\,B have similar temperatures. Based on the slope of the 
H-band peak, it is also justified to say that SONYC-NGC1333-42 is probably somewhat hotter than 2M1207\,B. There are some 
offsets in the flux level in the K-band, which may be explained by inaccuracies in the extinction and/or excess flux 
from a disk. Overall, the results of our analysis are confirmed.\footnote{The study by \citet{2012arXiv1201.3921P} 
includes three more young sources -- CT\,Cha, GQ\,Lup\,B, TWA5\,B. For these objects their comparison with models does 
not yield consistent temperatures and the fit with the model spectra is poor. Therefore we do not use them for the 
comparison, but these objects clearly pose a challenge for our understanding of substellar spectra.}

\begin{figure*}
\center
\includegraphics[width=4.0cm,angle=-90]{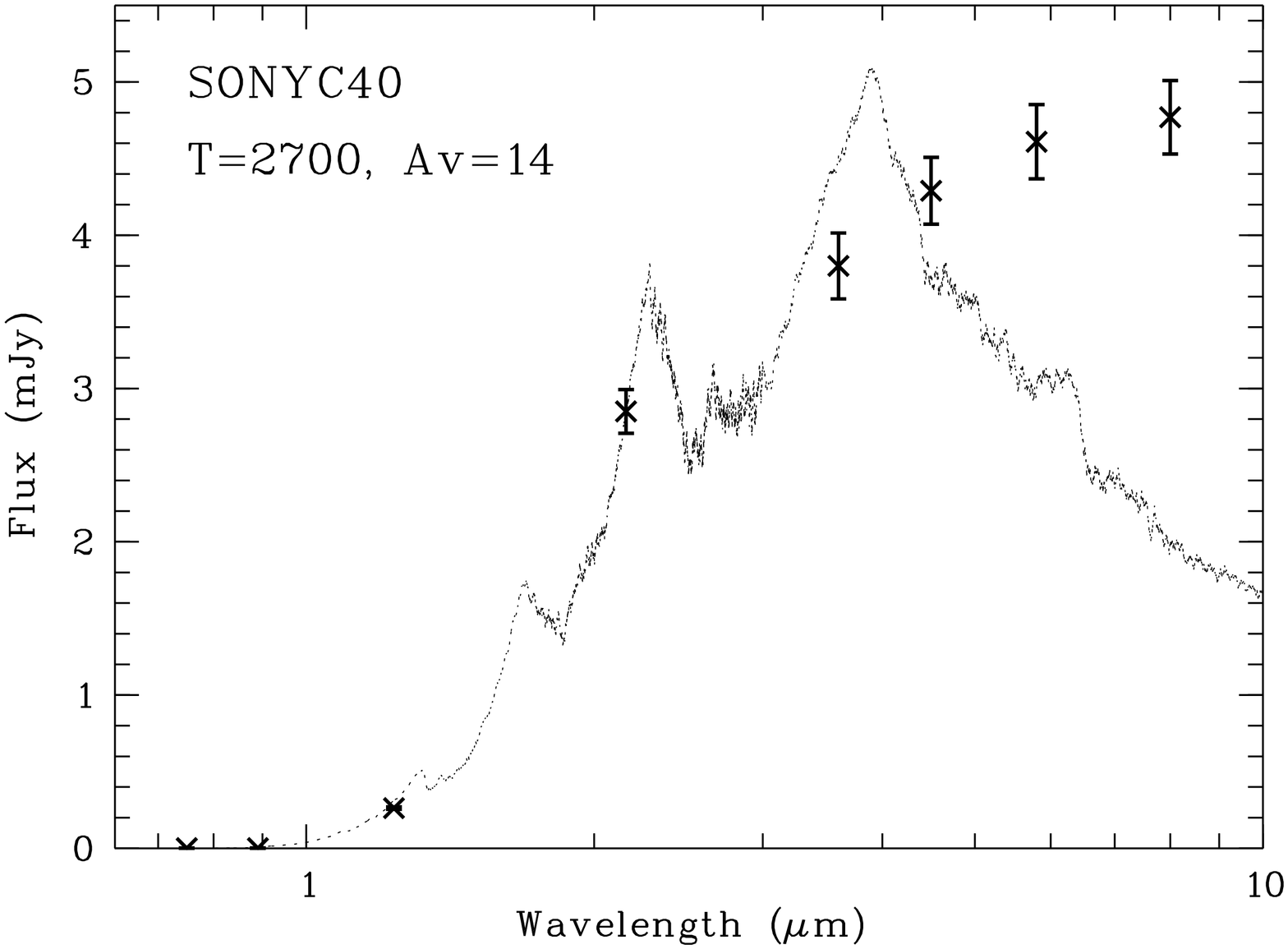} 
\includegraphics[width=4.0cm,angle=-90]{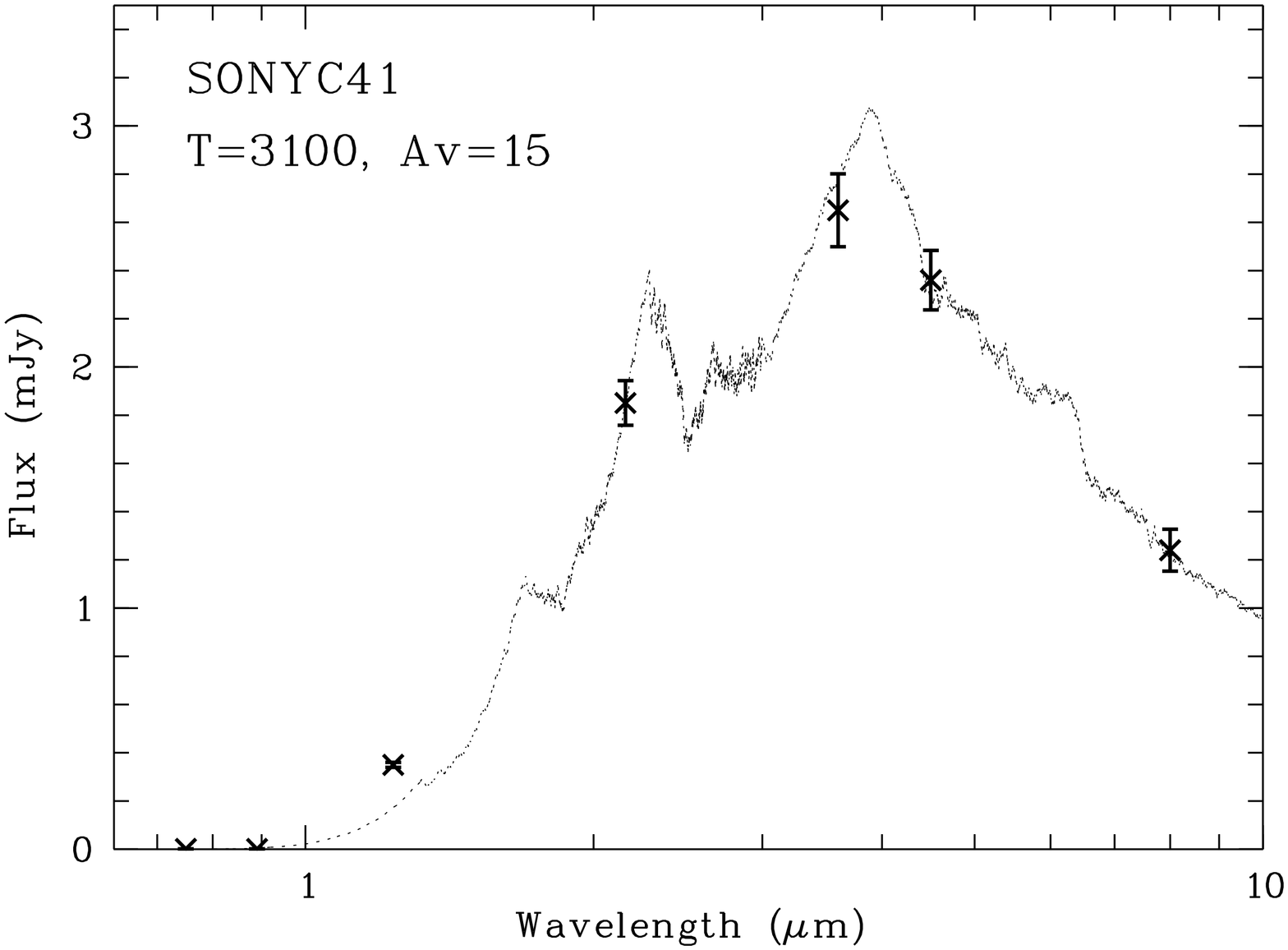} 
\includegraphics[width=4.0cm,angle=-90]{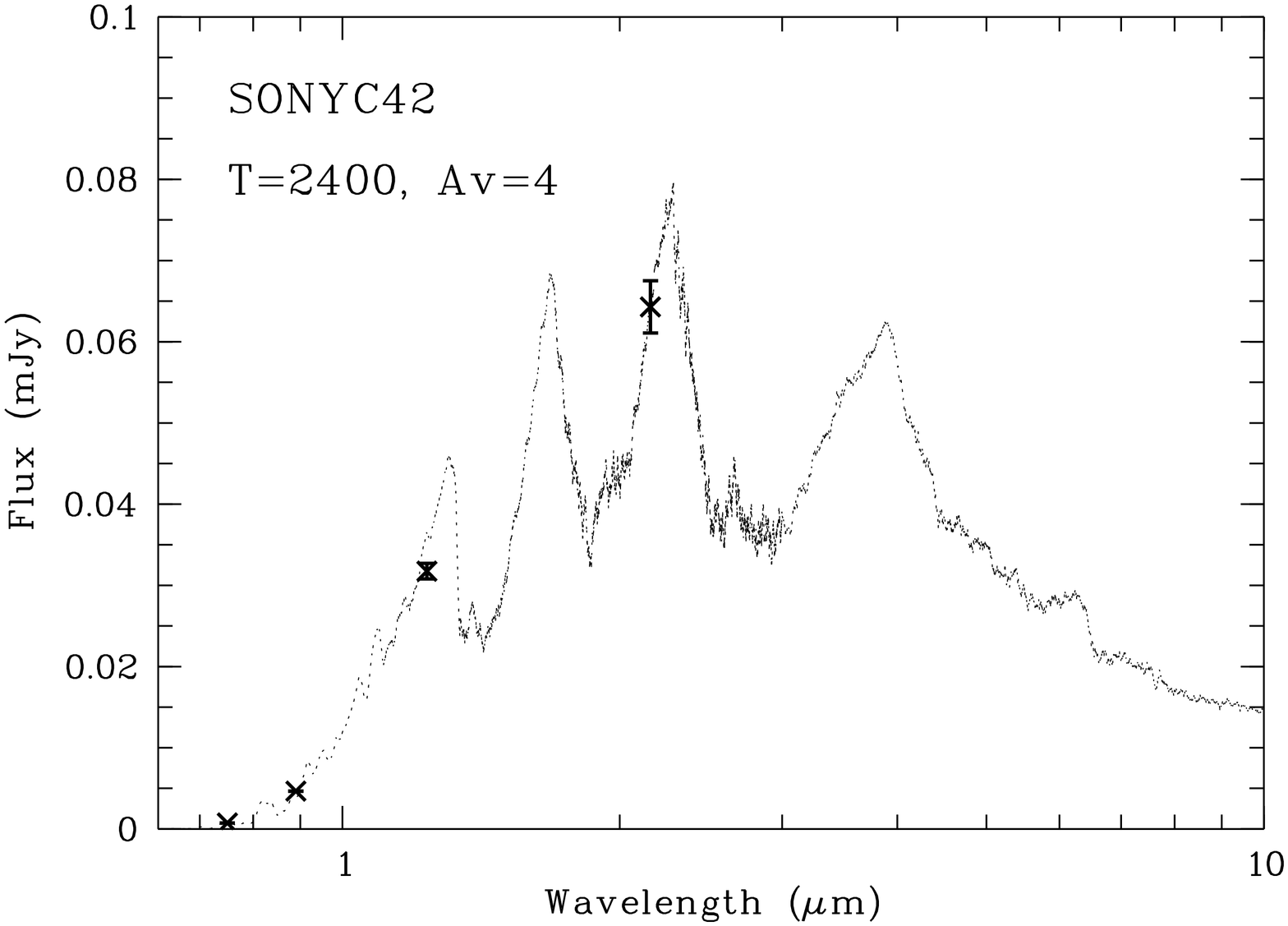} 
\caption{Photometric spectral flux distribution for three newly identified very low mass members of NGC1333 
in comparison with DUSTY model spectra for $T_{\mathrm{eff}}$ as given in Table \ref{t1}.
The model spectra have been dereddened using the photometric $A_V$ listed in Table \ref{t1}.
\label{f12}}
\end{figure*}

For the same three new SONYC objects, we compare the photometric fluxes from 0.8 to 8$\,\mu m$, as far as available,
with the model spectra, adopting $T_{\mathrm{eff}}$ and the photometric $A_V$ as listed in Table \ref{t1}, see Fig. \ref{f12}. 
Again, this comparison confirms the results from our analysis. For SONYC-NGC1333-40 mid-infrared excess is clearly visible and can
be attributed to the presence of a disk. As pointed out in SONYC-IV, spectroscopy covering the wide spectral range from 
the J- to the M-band promises to be the ideal tool to improve the characterisation of substellar objects.

\section{Discussion}
\label{s5}

With the new spectra presented in this paper we have enlarged the number of known very low mass members in NGC1333. 
In SONYC-IV we list 51 confirmed objects with spectral type M5 or later and/or effective temperature of 
3200\,K or lower (41 previously known and 10  new); this new campaign adds 7 new objects to this census. Based on 
the currently available isochrones for this mass domain, most of these 7 are cool enough to be considered to be 
brown dwarfs. In the following we will discuss the implications of the new results for the substellar population in 
NGC1333, with particular focus on the planemo domain.

\subsection{The distribution of spectral types and magnitudes}
\label{s50}

\begin{figure}
\center
\includegraphics[width=6cm,angle=-90]{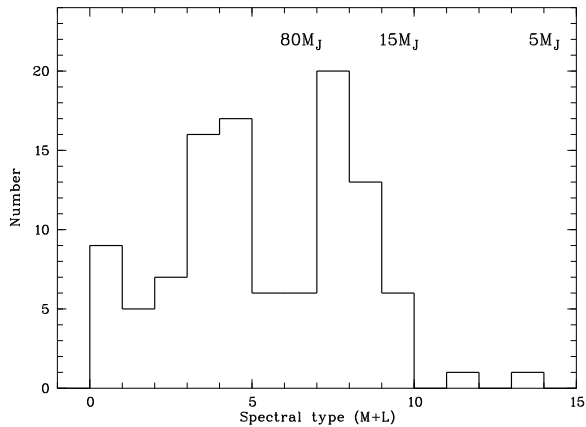} 
\caption{Distribution of spectral types for the very low mass population in NGC1333, including the previously
confirmed objects with $\ge$M5 listed in SONYC-IV, the new objects found in SONYC-IV and this paper, and the 
earlier type objects classified in \citet{2009AJ....137.4777W}. Note that Winston et al. additionally classify 
12 objects with spectral types earlier than M0. Approximate mass limits according to DUSTY/COND isochrones are 
given.
\label{f8}}
\end{figure}

We begin by showing the spectral type distribution for all spectroscopically classified M and L-type
objects in NGC1333 in Fig. \ref{f8}. This includes the very low mass objects listed in SONYC-IV, 
new objects identified in the same paper and this new study, plus earlier type objects classified in \citet{2009AJ....137.4777W}. 
Their spectroscopic survey is essentially complete down to the brown dwarf/star boundary (for $A_V<10$), and 
thus ideally complements our own studies, which are not complete above the substellar boundary.

For the stellar regime, the distribution of spectral types in NGC1333 resembles those found for other
regions (ChaI, IC348, Taurus, see \citealt{2007ApJS..173..104L}), with a peak around M4 and a increase in the 
frequency between M0 and M5 by about a factor of 3. For later spectral types, we observe a second peak around M7, 
which is not seen in these three other regions.

Another relevant feature in the context of this paper is the sharp drop of the numbers at M9. 
Based on the COND \citep{2003A&A...402..701B} and DUSTY \citep{2000ApJ...542..464C} isochrones, M9 corresponds to 
a mass around the Deuterium burning limit. Only two objects (SONYC-NGC1333-36 at $\sim$L3 and SONYC-NGC1333-42 at L1) 
have later spectral types. Thus, the number of observed planemos in this cluster is very low. This feature
will be further discussed in the following subsections.

\begin{figure}
\center
\includegraphics[width=6cm,angle=-90]{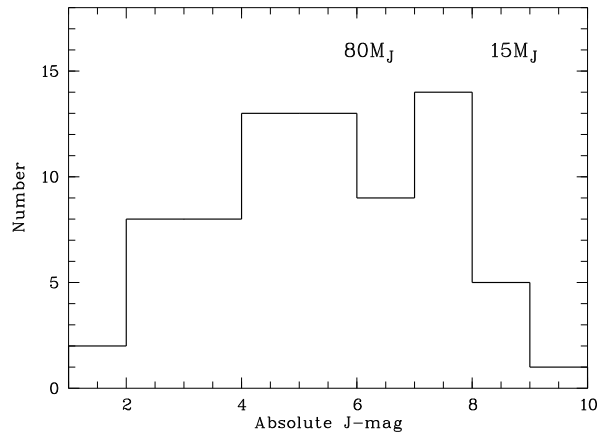} 
\caption{Distribution of absolute J-band magnitudes for the Class II in NGC1333 from \citet{2008ApJ...674..336G}
for all sources with $A_V<10$. Approximate mass limits according to DUSTY/COND isochrones are given.
\label{f4}}
\end{figure}

As an independent check of the distribution shown in Fig. \ref{f8}, we examine the histogram of absolute
J-band magnitudes for the population of Class II sources in NGC1333, as identified by \citet{2008ApJ...674..336G} 
based on Spitzer data. Of their 137 objects, 94 have photometry in 2MASS. We calculate $A_V$ from J- and K-band 
photometry, using the same prescription as in SONYC-IV, deredden the J-band photometry and subtract the distance 
modulus of 7.4 (assuming a distance of 300\,pc, see \citealt{2002A&A...387..117B}). We exclude the most reddened 
objects and consider only the 74 with $A_V<10$. In this extinction regime the distribution of J-band magnitudes 
does not show a dependence on $A_V$. 

Fig. \ref{f4} shows again a peak slightly above the substellar limit, which corresponds to the peak around
M4-M5 in the spectral type distribution. A second peak is visible in the substellar regime (for $M_J$ of 7-8),
but it is much less pronounced than the peak at M7 seen in Fig. \ref{f8}. In fact, within the statistical errors 
the magnitude distribution is consistent with a broad peak around $M_J = 5$ and a decline at fainter magnitudes, 
which agrees with previously published results for Cha-I and IC348 \citep{2007ApJS..173..104L}. Thus, the second 
peak in the spectral type distribution could be spurious. One possible explanation is a gap in completeness 
between the spectroscopic samples from \citet{2009AJ....137.4777W} and ours, causing us to miss objects
at $\sim$M6.

\subsection{The Hertzsprung-Russell diagram}
\label{s51}

\begin{figure}
\center
\includegraphics[width=6cm,angle=-90]{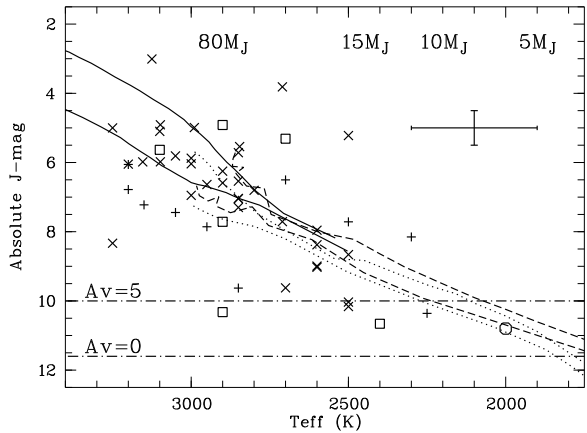} 
\caption{HRD for the very low mass population in NGC1333 (crosses -- previously known objects in the census
from SONYC-IV, plusses -- new objects found in SONYC-IV, squares -- new objects identified
in this paper). The circle shows the improved parameters for SONYC-NGC1333-36 after fitting the photometric SED,
see SONYC-IV.
Model isochrones (solid -- BCAH, dotted -- DUSTY, dashed -- COND) are shown for 1 (upper) and 5\,Myr (lower). The
dash-dotted lines show the limits of our spectroscopic survey for $A_V = 0$ and 5\,mag.
\label{f5}}
\end{figure}

In Fig. \ref{f5} we show the HRD for NGC1333, focusing on the very low mass objects with spectral types of 
M5 or later. Out of the 41 previously known sources listed in SONYC-IV (crosses), the 10 newly identified in 
SONYC-IV (plusses), and the 7 newly identified in this paper (squares), 51 have estimated effective temperatures 
and are plotted in the figure.

Instead of luminosity, we prefer to plot the absolute J-band magnitude to avoid additional errors from the
bolometric correction. The absolute J-band magnitudes are calculated in the same way as in Sect. \ref{s50}, including 
dereddening of the J-band data\footnote{As in SONYC-IV, we use an intrinsic J-K colour of 1.0, which is roughly 
applicable down to the planemo regime. For comparison, the latest type objects in $\sigma$\,Ori (7 objects) have 
an average J-K of 1.25 \citep{2007A&A...470..903C}. A shift in $J-K$ from 1.0 to 1.25 causes a shift of 1.36\,mag 
in $A_V$.} and subtraction of the distance modulus. The typical errorbar in the absolute J-band magnitude is $\pm 0.5$\,mag, 
which combines the errors in $A_V$ ($\pm 1$\,mag), distance ($\pm 20$\,pc), and photometry (up to $\pm 0.1$\,mag). 
The uncertainty in effective temperature, as outlined above, is $\pm 200$\,K. The approximate mass limits according 
to the DUSTY and COND isochrones, assuming an age of 1\,Myr, are given as well.

The general trend in this diagram is in good agreement with 3 sets of isochrones from the Lyon group: DUSTY, COND,
and BCAH \citep{1998A&A...337..403B}. Typical for HRDs of star forming regions, the objects show a large spread due 
to various effects \citep[e.g.][]{2011MNRAS.413L..56L}. However, only 7
objects (14\%) are inconsistent with the 1-5\,Myr isochrones within the errorbars. Some of them could be 
affected by ongoing accretion or strong magnetic activity (overluminosity) -- see \citet{2012MNRAS.419.1271S} for a 
discussion -- or grey extinction by an edge-on disk (underluminosity). Some might experience variability 
\citep{2012MNRAS.420.1495S}. In addition, there remains a small chance that an outlier is a fore- or background 
source: Perseus is known to have a complex substructure \citep{2006ApJ...638..293E,2006ApJ...643..932R}, i.e. there 
may be young objects at different distances. Moreover, in some cases the signal-to-noise in our spectra is not 
sufficient to confirm low gravity, i.e. we cannot completely exclude older objects in the field. Given these
caveats, the overall agreement with the isochrones is good. 

Only a very small number of 3 objects have temperature below 2500\,K -- SONYC-NGC1333-31, 36, and 42 --
(two of them also with spectral type later than M9, see Sect. \ref{s50}) and would have masses around or below the 
Deuterium burning limit. The plot demonstrates that the frequency of objects per bin in effective temperature drops 
significantly below 2500\,K. There are 19 objects with $2800 \le T_{\mathrm{eff}} < 3100$, 13 with 
$2500 \le T_{\mathrm{eff}} < 2800$, but only 3 with $2200 \le T_{\mathrm{eff}} < 2500$. This confirms the 
finding from the spectral type distribution.

The drop in the frequency of observed objects below 2500\,K and spectral type M9, already found in Sect. \ref{s50},
can either be explained by an actual lack of planemos in this cluster, or by biases in our photometric survey and 
the selection of objects for follow-up spectroscopy, which will be discussed in the following subsection. In addition, 
there are three more caveats:

1) Extinction might limit the completeness of our survey below 2500\,K. To test this, we have
overplotted with dash-dotted lines the J-band limit of our survey ($J=19$, see Sect. \ref{s2}) for $A_V=0$ and 
$A_V=5$. We should have been able to find objects with $T_{\mathrm{eff}}>2200$\,K and $A_V\le 5$. 
The clear majority of the confirmed very low mass objects have $A_V<5$. Also, $A_V\sim 5$ is the median
extinction for the sample of Class II objects published by \citet{2008ApJ...674..336G}, as far as their
2MASS photometry is available. Therefore we are unlikely to miss a significant number of planemos down 
to 2200\,K due to their high extinction. For extinctions well below $A_V$ of 5\,mag we should be able to 
find sources even cooler than 2200\,K (as proven by the detection of SONYC-NGC1333-36). 

2) The isochrones might not accurately reflect the luminosity vs. effective temperature relation in young star forming
regions. To explain a lack of planemos in Fig. \ref{f5}, the isochrones would have to turn downwards below 2500\,K, i.e
cooler objects need to be significantly fainter than expected based on these models. While this cannot be definitely 
excluded at the moment, the data for the eclipsing binary brown dwarf in Orion -- currently the only available empirical test 
for these isochrones  -- does not indicate such a trend \citep{2007ApJ...664.1154S}. The luminosities for its components are 
in agreement with the predictions from the isochrones, in fact, for the lower mass component ($M=0.036\,M_{\odot}$) 
the isochrone underestimates the luminosity, the reverse of what would be needed to explain the lack of planemos 
in NGC1333. Furthermore, it is difficult to conceive a significant departure from the smooth (almost linear) trend
in the isochrones in Fig. \ref{f6}, since this would require an additional and strong source of opacity in the J-band.
To our knowledge, this is not born out by the available data, for example, the same isochrones fit the optical and
near-infrared colours for the substellar population in $\sigma$\,Ori quite well down to the lowest masses \citep{2007A&A...470..903C}.

3) The effective temperatures -- mostly determined with the routines used in this paper, i.e. by fitting DUSTY
models to the near-infrared spectra -- could be systematically unreliable in a way that $T_{\mathrm{eff}}$ is 
overestimated for very cool sources. The study by \citet{2012arXiv1201.3921P}, already mentioned in Sect. \ref{s4},
does not indicate any such trend. Specifically, their comparison the DUSTY models 
yield effective temperatures below 2000\,K when compared with the spectra for 2M1207B, CT Cha B and GQ Lup B, 
three very young objects with L spectral types. This seems to work well even when only the H- and K-band spectra 
are considered, which is the case for our objects in NGC1333.

We conclude that these three caveats cannot be used to explain the lack of observed objects cooler than 2500\,K 
and later than M9.

\subsection{Survey completeness}
\label{s52}

In this subsection we will explore the possibility that our survey has missed a significant number of planemos
due to a lack of depth or biases in the colour cuts.

Our initial candidate selection was based on a (i, i-z) colour magnitude diagram (the IZ catalogue, see Sect. 
\ref{s2}). From the set of 196 candidates, more than half are now covered by follow-up spectroscopy. In particular,
27 out of 43 IZ candidates with $i>22$ and 9 out of 15 with $i>24$ have been verified spectroscopically. As illustrated
in Fig. \ref{f2}, the spectroscopy covers the entire colour-magnitude space of this selection down to the
faint end. This candidate sample contains 2 likely planemos (SONYC-NGC1333-31 is not found in this survey). 
Statistically, the remaining candidates which have not been verified yet are expected to contain one more.

This selection, however, might have missed planemos, because the depth of the IZ survey is not sufficient. 
As argued in SONYC-IV, this is a distinct possibility; while the nominal completeness limit is $i=24.8$, 
it might be lower in some parts of the cluster with strong background emission.

To overcome the depth limitation of the IZ survey, we have used the JKS catalogue, which contains all objects with J- and K-band
photometry from our own survey and Spitzer photometry in the first two IRAC bands at 3.6 and 4.5$\,\mu m$ (see Sect. \ref{s2}).
As illustrated in Fig. \ref{f3}, this survey is much deeper and extends significantly beyond the limiting magnitude for
follow-up spectroscopy ($J=19$). It should be complete down to $J=20.8$. The near-infrared images do not show strong background
structure due to the cloud, i.e. the completeness should be uniform across the cluster. The JKS catalogue contains about
two thirds of the confirmed very low mass objects in NGC1333; most of the ones that are not in this catalogue are 
relatively bright and therefore affected by saturation in our deep near-infrared images. This catalogue contains
2 likely planemos (SONYC-NGC1333-42 does not have a Spitzer counterpart).

The colour-magnitude diagram from this catalogue is shown in Fig. \ref{f3}. Our spectroscopic 
follow-up covers a large fraction of this diagram to the $J=19$ limit. According
to the COND and DUSTY 1\,Myr isochrones, this should correspond to a mass of 0.003-0.004$\,M_{\odot}$ for $A_V=0$
and 0.006-0.008$\,M_{\odot}$ for $A_V=5$. For $15<J<17$ the entire colour range in $I1-I2$ has been probed by 
spectroscopy; it turns out that most of the confirmed objects in this magnitude range (11/13) have $I1-I2>0.2$. As 
the IRAC colour can be affected by excess emission due to a disk, this simply might reflect the high disk fraction 
in NGC1333 (see SONYC-IV). For $17<J<19$ we verified about three quarters of the objects with $I1-I2>0.3$. 
Thus, in these colour domains the spectroscopic survey is sufficiently complete to rule out the existence
of more than 1-2 additional planemos.

\begin{figure}
\center
\includegraphics[width=6cm,angle=-90]{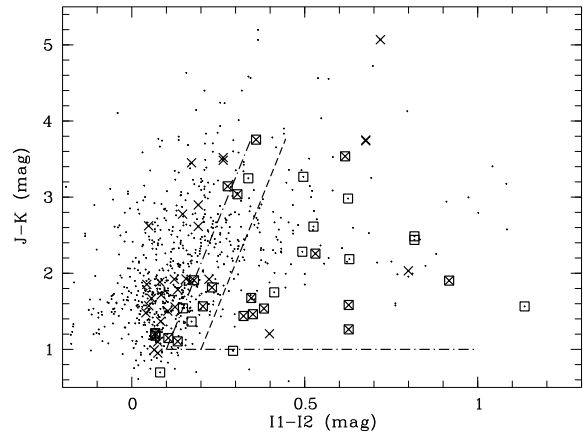} 
\caption{Colour-colour plot for the JKS catalogue (see Sect. \ref{s2}). Marked are objects for which
we have taken spectra (crosses) and confirmed very low mass members of NGC1333 (squares). The dash-dotted lines
mark the colour space where brown dwarfs and planemos are expected to be found. The dashed line shows the 
reddening path for the average planemo.
\label{f7}}
\end{figure}

To further analyse the colour space from the JKS catalogue, we show a (J-K, I1-I2) colour-colour diagram 
in Fig. \ref{f7}. In comparison with other colour combinations in this catalogue, this plot shows a 
relatively clear distinction between objects confirmed as very low mass cluster members and those rejected.
All confirmed objects are marked with squares, all objects verified with spectroscopy with crosses. 
For this plot, we estimate the expected position of planemos based on the photometry of known young ultracool 
objects: In the $\sigma$\,Ori cluster \citep{2008ApJ...672L..49S} the average colour of 16 M9-L5-type objects 
is $I1-I2 = 0.2$, albeit with a spread of $\pm 0.2$\,mag due to photometric uncertainties. Among 10 young 
field dwarfs with L0-L4 spectral type, the average colour is $I1-I2 = 0.17$ with a spread of 0.05\,mag \citep{2009ApJ...703..399L}. 
Based on these two samples, we adopt $I1-I2 = 0.2$ as typical intrinsic colour of planemos.

The dashed line in Fig. \ref{f7} shows the reddening path for $A_V = 0-15$ starting with this intrinsic 
colour. Shifting this line by 0.1\,mag to the left is in good agreement with the left boundary for most of the confirmed 
very low mass objects in NGC1333 (dash-dotted line). This line combined with a limit of $J-K>1.0$ is adopted as 
the colour space where brown dwarfs and planemos are expected.

In this regime we have 250 objects out of which 103 have $16<J<19$ and are thus sufficiently faint to be
planemos ($J=16$ corresponds to $\sim 0.015\,M_{\odot}$ for zero extinction in COND/DUSTY isochrones) and 
sufficiently bright to be accessible for spectroscopy. About one quarter of them has been observed spectroscopically. 
Thus, the catalogue could contain a few additional planemos -- up to 6, statistically, which would yield a total
of up to 8. This analysis shows that one particular bias in the selection of candidates, the cutoff at $I1-I2=0.3$ 
which was necessitated by the large amounts of contaminating objects at bluer colours, might affect our ability 
to find planemos. 

We should not miss any significant number of planemos with enhanced mid-infrared colours due to the 
presence of a disk. If the disk fraction for planemos is as high as for brown dwarfs, as shown for the 
$\sigma$\,Ori cluster \citep{2008ApJ...672L..49S}, excess emission should be expected for about half of 
the planemos in NGC1333 (see SONYC-IV for more information on disk fractions). These objects should be
located to the right of the dashed line. In this area we find 39 objects from which about half have been
verified spectroscopically (including the two planemos). This implies that we might miss about two more 
objects in this colour domain. Assuming a disk fraction of around $>50$\%, this again gives a total number
of up to 8 planemos.

The JKS catalogue can further be used to probe the regime $J>19$ which is not accessible for spectroscopy.
The JKS catalogue contains 292 objects with $J>19$, 124 in the colour space expected for planemos (dash-dotted 
lines in Fig. \ref{f7}). However,
this sample is dominated by strongly reddened objects with high $J-K$ colour. Only about 
30 of them have $J-K$ colours consistent with $A_V < 5$ and only 7 are below $A_V = 3$. Given that 
two thirds of the confirmed brown dwarfs have an optical extinction below 5\,mag and about half below 
3\,mag, this basically rules out that the candidates too faint for spectroscopy down to $J\sim 21$
contain a sizable population of planemos. According to the COND and DUSTY isochrones, this limit 
corresponds to object masses of 0.002-0.003$\,M_{\odot}$ (for $A_V$ of 0-5 and age of 1\,Myr).

\subsection{The frequency of planemos}
\label{s53}

In the following we will use the survey results in NGC1333 to give a quantitative constraint
on the frequency of planemos which can be compared with other star forming regions. To do this,
we estimate the fraction of brown dwarfs which have a mass in the planemo domain, $f_P = N_P / N_{\mathrm{BD}}$. 
This quantity serves as a proxy for the shape of the mass function in the substellar regime.
Here we treat the population of planemos as a subsample of the population of brown dwarfs.

Our survey in NGC1333 yields only 3 objects with effective temperatures below 2500\,K or 
spectral types later than M9. According to the currently available 1-5\,Myr isochrones these objects 
have masses below the Deuterium burning limit of 0.015$\,M_{\odot}$ and are thus good candidates
for planemos. As outlined above, we might miss about 6 more planemos due to our cutoff in the
mid-infrared colour.

The total number of brown dwarfs can be calculated as follows: Adding the objects listed in SONYC-IV 
and this paper, there are 58 with spectral type M5 or later or effective temperatures
of 3200\,K or below. From these around 10-20 are likely to be above the substellar boundary, which
leaves 38-48 brown dwarfs. Following the same statistical arguments that we have used in Sect. 4.1. 
of SONYC-IV we estimate that we might miss up to 13 more brown dwarfs, plus 6 planemos (see above), 
i.e. the total is 57-67.  

Thus, for NGC1333 we estimate that $f_P$ is 12-14\%. Just counting the number of confirmed
objects, the ratio becomes 6-8\%. This includes all planemos down to masses of 0.006-0.008$\,M_{\odot}$ 
(the mass limit for $J=19$ and $A_V=5$).

This value can be compared with other star forming regions. This comparison comes with three caveats:
1) Only $\sigma$\,Ori and UpSco have been surveyed with a depth comparable to our survey in NGC1333, 
i.e. all other ratios should be considered lower limits.
2) While the general survey strategy is similar in all cases (candidate selection based
on optical/near-infrared photometry plus follow-up spectroscopy), the details of the strategy 
differ from region to region, which might introduce unknown biases. To be conservative, we focus
in the following mostly on samples with spectroscopic confirmation. 3) Most of the studies listed 
below do not make an attempt to estimate the {\it total} substellar population, by taking into 
account possible survey biases.

We note that all existing surveys use the COND/DUSTY isochrones for mass estimates, as in our
paper. In the case these isochrones turn out to be systematically off, the results of the comparison 
should still be valid and merely shifted to different mass regimes. 

{\it $\sigma$\,Ori:} The summary paper by \citet{2007A&A...470..903C} lists 46 confirmed brown dwarfs,
out of which 12 are expected to be in the planemo regime, i.e. $f_P \sim 26$\%. Not all of these objects
are spectroscopically confirmed, but the contamination in this cluster is generally low due to the negligible
extinction. The surveys in this region are similarly deep as in NGC1333, i.e. the numbers should be comparable.

{\it Upper Scorpius:} Based on photometry and proper motions from UKIDSS, the total number of brown dwarfs in
this region is 68 in the UKIDSS covered area, out of which 49 are in the northern area where follow-up 
spectroscopy is available \citep{2011MNRAS.418.1231D}. The currently ongoing spectroscopic verification shows
that essentially all these objects are indeed members of UpSco \citep{2008MNRAS.383.1385L,2011A&A...527A..24L}. 
Converting the photometry to masses, this sample should cover the mass range down to 0.01$\,M_{\odot}$ and contains 
about half a dozen objects below $0.015\,M_{\odot}$. A deeper survey by \citet{2011MNRAS.418.2604L} does not 
find new members in the planemo domain.\footnote{Here we neglect the 5 T-dwarf candidates identified based on imaging in 
Methane-sensitive narrow-band filters without spectroscopic or proper motion confirmation.} This yields $f_P \sim 10$\%.

{\it $\rho$-Ophiuchus:} This region is affected by very high and variable extinction, which strongly hampers
any survey. The existing census is not nearly complete below 0.03$\,M_{\odot}$. So far, there are around 
40 confirmed objects with spectral types later or equal M6 from 
\citet{2008hsf2.book..351W,2011ApJ...726...23G,2010ApJ...709L.158M,2012ApJ...744..134M,2010A&A...515A..75A,2012arXiv1201.1912A}.
Out of these, 8 might be in the planemo regime based on spectral type and effective temperature, which results in 
$f_P \sim 20$\%.

{\it Chamaeleon-I:} The census by \citet{2007ApJS..173..104L,2008ApJ...684..654L} contains 35 objects with spectral 
type of M6 or later which are to be considered brown dwarfs. 5 of these have estimated masses below 0.015$\,M_{\odot}$, 
i.e. $f_p \sim 14$\%. We note that our own deep survey work in Cha-I has not revealed any additional planemos yet 
down to 0.008$\,M_{\odot}$ \citep{2011ApJ...732...86M}.

Taking into account the caveats listed above, the values for $f_P$ in the various regions are consistent with 
each other and in the range between 10 and 20\%; only $\sigma$\,Ori might have a slightly larger planemo fraction. 
Focusing on the confirmed objects, NGC1333 has a relatively low planemo fraction compared with other star forming
areas. 

We can also estimate the fraction of planemos with respect to the population of young stars, $g_P = N_P / N_{\mathrm{S}}$. 
For NGC1333 we have recently determined that the star-vs-brown dwarf ratio, defined as 
$R_1 = N(0.08-1.0\,M_{\odot}/N(0.03-0.08\,M_{\odot})$, is $2.3\pm 0.5$, for most other regions this
ratio is larger. Taking this into account, for NGC1333 the fraction of planemos with respect to the total cluster 
population is in the range of 2-3\%. For the other regions where this type of census is available this number becomes 
2-5\%, i.e. stars are 20-50 times more frequent than planemos. 
This is consistent with the planemo fraction of $<10$\% \citep{2005MNRAS.361..211L} and 
1-14\% \citep{2006MNRAS.373L..60L} derived for the Orion Nebulae Cluster (ONC), albeit for slightly 
different mass limits. Thus, the number of planemos is insignificant compared 
with the total population of stars and brown dwarfs. The planemo contribution to the mass budget of young clusters is 
negligible (well below 1\%).

\subsection{The mass spectrum}
\label{s54}

With the available database of spectroscopically confirmed objects in NGC1333, we are in the
position to construct the Initial Mass Function (IMF) in the low-mass regime. We choose to express
the IMF as a mass spectrum $dN/dM \propto M^{-\alpha}$. To do that, we use the same spectroscopic 
samples as in Sect. \ref{s50}, combining the confirmed low-mass stars from \citet{2009AJ....137.4777W}
with the census of very low mass objects. We select all M and L dwarfs, in total 107 objects,
out of which 100 have an estimate for the effective temperature, either from our survey or from
\citet{2009AJ....137.4777W}. 

Masses were derived for this sample by comparing the temperatures with theoretical evolutionary
tracks. Temperatures do not significantly depend on age in the 1-10\,Myr phase, as opposed to luminosities
and magnitudes, and thus provide a mass estimate that is insensitive to age spread.
We use the 1\,Myr isochrone from the BCAH models for $M\ge 0.1\,M_{\odot}$ and the 
1\,Myr isochrone from the DUSTY models for $M<0.1\,M_{\odot}$. For each object with a given 
$T_{\mathrm{eff}}$, we selected the isochrone datapoints within $\pm 200$\,K around $T_{\mathrm{eff}}$,
fitted this section either linearly (if there are less than 5 datapoints) or with a second order
polynomial, and used that function to calculate the mass corresponding to $T_{\mathrm{eff}}$. All
masses are tied to this isochrone and should not be compared with values derived using other models.

\begin{figure}
\center
\includegraphics[width=6cm,angle=-90]{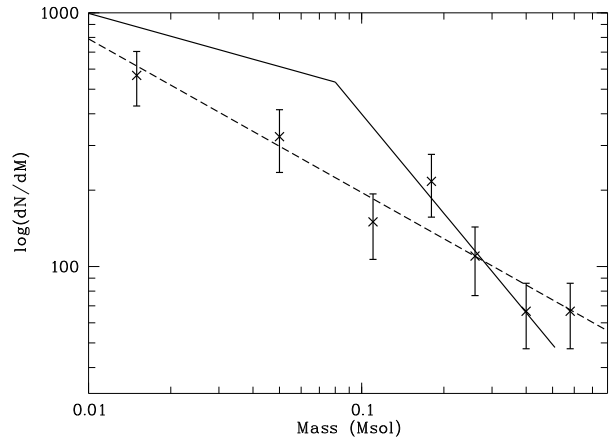} 
\caption{Mass function in NGC1333 based on the spectroscopically confirmed members with measured
effective temperatures. The binsizes were chosen to achieve approximately equal error bars. The
solid lines show the Kroupa mass function ($\alpha = 1.3$ for $0.08<M<0.5\,M_{\odot}$, $\alpha = 0.3$
for $M<0.08\,M_{\odot}$). The dashed line shows the best fit to the data ($\alpha = 0.6$).
\label{f9}}
\end{figure}

The mass spectrum was calculated in a way to achieve similar statistical uncertainties in each
mass bin, i.e. with varying bin size. For $M>0.7\,M_{\odot}$ the sample sizes are too small for a 
meaningful analysis, the 10 objects above this limit were excluded. The lowest mass bin includes
all objects with $M<0.03\,M_{\odot}$. In Fig. \ref{f9} we plot our 
result. The datapoints in this diagram are consistent with $dN/dM \propto M^{-\alpha}$ 
with $\alpha = 0.61$, overplotted as dashed line. For the substellar regime alone, the best-fitting 
slope is $\alpha = 0.46$. For the stellar regime above 0.1$\,M_{\odot}$ the slope becomes $\alpha = 1.0$.
The uncertainty in the slope is in the range of 0.1.

To verify how the uncertainties in the effective temperatures ($\pm 200$\,K) affect 
these results, we carried out the following test: For half the objects, randomly selected, we
increased $T_{\mathrm{eff}}$ by 200\,K, for the other half we reduced it by 200\,K. This effectively
scrambles the values within the errorbars. Then the mass spectrum was calculated again using the same 
procedure. This yields $\alpha = 0.66$ for the entire sample and 0.55 for the brown dwarfs, slightly 
higher as before but well within the errorbars. This test also shows that the peak around 
0.2$\,M_{\odot}$ seen in the figure is spurious.

Also shown is the Kroupa IMF, an empirical description of the mass spectrum as a segmented power law,
as a solid line \citep{2001MNRAS.322..231K}. This function follows $\alpha = 1.3\pm 0.5$ in the stellar regime 
up to 0.5$\,M_{\odot}$ and $\alpha = 0.3\pm 0.7$ in the substellar domain. Within the uncertainties, the 
Kroupa IMF is consistent with the mass spectrum in NGC1333.

The slope derived here is broadly consistent with values derived for most other star forming regions and young clusters. 
Some examples: For the very low mass objects in $\sigma$\,Ori, values of 0.6-0.7 have been published in the literature
\citep{2007A&A...470..903C,2011ApJ...743...64B}. For the domain from 0.03 to 0.6\,$M_{\odot}$ the cluster Blanco 1
has a mass function with $\alpha = 0.69 \pm 0.15$ \citep{2007A&A...471..499M}. For Upper Scorpius and the mass
regime 0.01-0.3$\,M_{\odot}$, a value of $\alpha = 0.6 \pm 0.1$ has been published \citep{2007MNRAS.374..372L}. 
For the Pleiades $\alpha$ is found to be in the range of 0.6-1.0 in the low-mass regime 
\citep[e.g.][and references therein]{2003A&A...400..891M}. For the $\alpha$\,Per cluster, the slope is 
$0.59 \pm 0.05$ \citep{2002A&A...395..813B}. In Collinder\,69 the slope has been found to be between 0.2 and
0.4 for the low-mass regime \citep{2011A&A...536A..63B}. 

To our knowledge, the only young region where a significantly different slope of the mass spectrum has been determined 
is the Orion Nebulae Cluster. In a recent paper, \citet{2011arXiv1112.2711D} find $\Gamma$ of $-1.1$ to $-3.1$ for
the substellar regime, which corresponds to a negative $\alpha$ ($\alpha = \Gamma + 1$). Similar results have been found in
previous studies in the ONC \citep[e.g.][]{2002ApJ...573..366M}. These numbers appear to be anomalous compared 
with the other regions mentioned above.

Apart from the overall slope, we can examine the shape of the mass spectrum in NGC1333. In equal sized bins, the 
numbers per bin increase continuously with decreasing mass, with the exception of a slight minimum around 
0.1-0.2$\,M_{\odot}$. However, the test with the modified temperatures (see above) indicates that this minimum 
is spurious, the data is indeed consistent with a monotonically rising histogram. The same result has been obtained 
for Upper Scorpius and $\sigma$\,Ori, while some other regions show a peak in the mass spectrum around 
0.1-0.2$\,M_{\odot}$ and a smaller slope for lower masses (IC348, Cha-I, $\rho$-Oph, 
\citealt{2007ApJS..173..104L,2012arXiv1201.1912A}). One consequence of this difference is a smaller 
star-to-brown-dwarf ratio (i.e. a large number of brown dwarfs relative to stars) in regions with monotonically 
rising mass spectrum -- as it is indeed found for NGC1333. As discussed in SONYC-IV, a comparison of these
ratios possibly indicates environmental differences in the formation of brown dwarfs. 

The focus of this study is the planetary mass domain. Assuming a monotonic continuation of the power law shown in Fig. 
\ref{f9} with $\alpha = 0.6\pm 0.1$, we would expect $8\pm 3$ objects with $0.005<M<0.015\,M_{\odot}$, i.e. in the planemo 
domain and detectable in our survey. For comparison, in Sect. \ref{s52} we derive an upper limit of 8 planemos in this
mass regime from our survey, taking into account the incompleteness in the spectroscopic follow-up of our candidates.
Thus, the data is consistent with a monotonic power law mass spectrum across the Deuterium burning limit, but a smaller 
slope is possible as well. Our survey rules out that the slope of the mass spectrum in NGC1333 increases in the planemo 
regime.

Again this is in agreement with previous findings for other regions. In $\sigma$\,Ori the census of substellar
objects is well approximated by a monotonic slope down to planetary masses 
\citep{2007A&A...470..903C,2011ApJ...743...64B,2012arXiv1205.4950P}, with a possible turnover (i.e. a lower $\alpha$) 
in the mass spectrum below 0.006$\,M_{\odot}$. For Upper Scorpius, the results by \citet{2011MNRAS.418.2604L} favour 
a turndown below 0.01$\,M_{\odot}$. We note that \citet{2010ApJ...719..550M} show a mass function of $\rho$-Ophiuchus 
covering the entire planemo regime based on a photometrically selected candidate sample, which resembles the one in 
$\sigma$\,Ori. Taken together, it seems safe to conclude that the slope of the mass function in the planemo domain 
is $\alpha \la 0.6$.

We note that $\alpha \la 0.6$ corresponds to a planemo fraction as defined in Sect. \ref{s53} of $f_P \la 20$\% 
(if the entire planemo domain is considered) or $f_P \la 30$\% (with an lower mass cutoff at 0.005\,$M_{\odot}$).
Indeed, no regions shows a larger planemo fraction, most surveys indicate significantly lower values, including
our own work in NGC1333. 

Thus, the current census of star forming regions provides support for the idea that planemos are an extension of the
population of stars and brown dwarfs and form through the same mechanisms. The observations indicate that free-floating 
objects with planetary masses exist down to at least 0.006$\,M_{\odot}$, but, as pointed out in Sect. \ref{s53}, their 
numbers and their mass budget are insignificant compared with the total population of stars and brown dwarfs.

Some theoretical estimates for the lower mass limit in the star formation 
process -- the opacity limit for fragmentation -- are 0.007 \citep{1976MNRAS.176..367L} or 0.01$\,M_{\odot}$ 
\citep{1977ApJ...214..152S} from analytical arguments and $<0.001$ \citep{2001ApJ...551L.167B} or 0.003-0.009$\,M_{\odot}$
\citep{2005MNRAS.363..363B} from numerical simulations. Except for the lowest-mass values, all these estimates
are consistent with the current observational picture.

\citet{2011Natur.473..349S} recently determined the slope of the mass spectrum based on microlensing surveys,
and find $1.3 \pm ^{0.3}_{0.4}$ for $M<0.01\,M{\odot}$, significantly higher than for brown dwarfs (0.49 in their
work), indicating that Jupiter-mass objects are almost twice as common as stars. Their estimate includes 
planetary-mass objects without host stars (i.e. free-floating) and those on wide orbits, but they argue that 
the majority of them are indeed unbound (but see \citet{2012A&A...541A.133Q} for a criticism of this claim).
This is in clear disagreement with the results from the surveys in star forming regions, where the number of 
planemos per star is much smaller (see Sect. \ref{s53}) and the slope in the mass spectrum is not rising (see above). 
If the microlensing result holds, the additional planemos would have to be in the mass regime $<0.005\,M_{\odot}$ 
which is not sufficiently probed by the cluster surveys. This would imply a break in the slope of the mass spectrum 
in the planemo domain, favouring a scenario where the Jupiter-mass objects form in a different way than the planemos 
and brown dwarfs seen in the star forming regions -- as indeed concluded by \citet{2011Natur.473..349S}. Larger 
samples of short-period microlensing events and deeper surveys in star forming regions are critical to understand 
if they are in fact different populations or not.

In another recent paper, \citet{2012MNRAS.tmp.2972S} estimate that there may be up to $10^{5}$ objects with masses 
between $10^{-8}$ and $10^{-2}\,M_{\odot}$ (called 'nomads' in their paper) per main-sequence star. 
To arrive at this number, they use the slope of the mass spectrum in the brown dwarf/planemo regime as a constraint. 
Their upper limit of $10^5$ can only be obtained by using $\alpha = 2$. As shown above, this is inconsistent 
with the surveys in star forming regions. For a more realistic value of $\alpha = 0.5$, the result given
in their paper is $\ga 1$ nomads per main-sequence star. Given the enormous difficulties in extrapolating the 
mass function over 6 orders of magnitude in mass and different modes of formation, these estimates remain
highly speculative. 

\section{Conclusions}
\label{s6}

We present new spectroscopy for 19 very faint candidate members in the young cluster NGC1333. 7 of them are confirmed
as very low mass objects, most or all of them should be brown dwarfs. Based on the combined spectroscopic follow up 
from SONYC-I, SONYC-IV and this paper, we are able to put limits on the IMF and the number of free-floating planetary-mass
objects (planemos, $M<0.015\,M_{\odot}$) in this cluster. The mass spectrum is in good agreement with a monotonically
rising power law $dN/dM \propto M{^{-\alpha}}$, with $\alpha = 0.6\pm 0.1$ for $M<0.6\,M_{\odot}$. The overall slope
is in agreement with all previous studies in other regions, with the ONC as only exception.

Among the substellar population in NGC1333 are 3 with estimated masses in the planemo domain. Taking into account
the incompleteness of the spectroscopic follow-up in our survey, we estimate that the cluster could harbour a total of 
8 planemos down to masses of $\sim 0.005\,M_{\odot}$. This translates into a planemo fraction (the fraction of brown 
dwarfs in the planetary mass domain) of 12-14\%. This is comparable to the planemo fractions determined for other 
regions. We estimate that there are 20-50 times more stars than planemos in young clusters. Based on the constraints on the 
number of planemos in star forming regions, the slope of the mass spectrum is $\la 0.6$ below 0.015$\,M_{\odot}$, i.e. it 
does not increase in the planemo domain. These findings support the idea that planemos form through the same mechanisms
as young stars and brown dwarfs.
 
\acknowledgments
The authors would like to thank the Subaru staff for the assistance during the observations and 
their preparation. We thank Jenny Patience for making her spectra available to us. The careful review
by an anonymous referee helped us to improve the manuscript.
AS acknowledges financial support the grant 10/RFP/AST2780 from the Science Foundation 
Ireland. The research was supported in large part by grants from the Natural Sciences and Engineering 
Research Council (NSERC) of Canada to RJ. RJ's work was also supported in part by the Radcliffe 
Institute for Advanced Study at Harvard University.

\bibliography{aleksbib}

\end{document}